\documentclass[%
pre,
aps,
aip,%
amsmath,amssymb,floatfix,
reprint,%
twocolumn,%
jcp,%
longbibliography,
reprint,%
]{revtex4-1} 

\usepackage{amsmath,amsthm,latexsym,amssymb,amsfonts,epsfig}

    \usepackage[english]{babel}
    \usepackage[utf8]{inputenc}			
    \usepackage{graphicx, texdraw}   
    \usepackage{bm}

    \usepackage{lmodern}
    \usepackage{mathtools}
    \usepackage{mathrsfs} 

    \usepackage[left=1.3cm, right=1.3cm, top=2cm, bottom=2cm]{geometry}     

    \usepackage{enumerate}

    \usepackage[colorlinks=true,citecolor=cyan, linkcolor=blue]{hyperref}

    \usepackage{csquotes} 

    \usepackage[usenames, dvipsnames]{color}

    \usepackage{amsmath}
    \usepackage{amsfonts}
    \usepackage{amssymb}
    \usepackage{braket}

    \usepackage[normalem]{ulem} 

\newcommand{\vect}[1]{\boldsymbol{#1}}

\newcommand{\boldNabla}{\boldsymbol{\nabla}}

\newcommand{\boldSigma}{\boldsymbol{\sigma}}

\newcommand{\kS}{\kappa_\mathrm{S}}
\newcommand{\kA}{\kappa_\mathrm{A}}
\newcommand{\kB}{\kappa_\mathrm{B}}
\newcommand{\EB}{E_\mathrm{B}}

\newcommand{\alphaB}{\alpha_\mathrm{B}}

\newcommand{\G}{\mathcal{G}}
\newcommand{\R}{\vect{r}}
\newcommand{\Intd}{\mathrm{d }}

\newcommand{\F}{\vect{F}}

\newcommand{\X}{\vect{x}}

\newcommand{\bigO}{\mathcal{O}}

\newcommand{\Faxen}{Fax\'{e}n}

\newcommand{\eR}{\vect{e}_r}

\newcommand{\eThe}{\vect{e}_{\theta}}

\newcommand{\vStok}{\vect{v}^{\mathrm{S}}}
\newcommand{\vStokcom}{{v}^{\mathrm{S}}}
\newcommand{\pStok}{p^\mathrm{S}}

\newcommand{\xOne}{\vect{x}_1}
\newcommand{\xTwo}{\vect{x}_2}

\newcommand{\vecD}{\vect{d}}
\newcommand{\infSum}{\sum_{n=0}^{\infty}}
\newcommand{\infSumOne}{\sum_{n=1}^{\infty}}

    \usepackage{epstopdf}

\usepackage{type1ec} %

\begin{document}
\title{Creeping motion of a solid particle inside a spherical elastic cavity}

\author{Abdallah Daddi-Moussa-Ider}
\email{ider@thphy.uni-duesseldorf.de}

\affiliation
{Institut f\"{u}r Theoretische Physik II: Weiche Materie, Heinrich-Heine-Universit\"{a}t D\"{u}sseldorf, Universit\"{a}tsstra\ss e 1, 40225 D\"{u}sseldorf, Germany}

\author{Hartmut Löwen}

\affiliation
{Institut f\"{u}r Theoretische Physik II: Weiche Materie, Heinrich-Heine-Universit\"{a}t D\"{u}sseldorf, Universit\"{a}tsstra\ss e 1, 40225 D\"{u}sseldorf, Germany}

\author{Stephan Gekle}
\affiliation
{Biofluid Simulation and Modeling, Theoretische Physik VI, Universit\"at Bayreuth, Universit\"{a}tsstra{\ss}e 30, 95440 Bayreuth, Germany}

\date{\today}


\begin{abstract}
On the basis of the linear hydrodynamic equations, we present an analytical theory for the low-Reynolds-number motion of a solid particle moving inside a larger spherical elastic cavity which can be seen as a model system for a fluid vesicle.
In the particular situation where the particle is concentric with the cavity, we use the stream function technique to find exact analytical solutions of the fluid motion equations on both sides of the elastic cavity.
In this particular situation, we find that the solution of the hydrodynamic equations is solely determined by membrane shear properties and that bending does not play a role.
For an arbitrary position of the solid particle within the spherical cavity, we employ the image solution technique to compute the axisymmetric flow field induced by a point force (Stokeslet).
We then obtain analytical expressions of the leading order mobility function describing the fluid-mediated hydrodynamic interactions between the particle and confining elastic cavity.
In the quasi-steady limit of vanishing frequency, we find that the particle self-mobility function is higher than that predicted inside a rigid no-slip cavity.
Considering the cavity motion, we find that the pair-mobility function is determined only by membrane shear properties.
Our analytical predictions are supplemented and validated by fully-resolved boundary integral simulations where a very good agreement is obtained over the whole range of applied forcing frequencies.

\end{abstract}
\maketitle


\section{Introduction}

Transport phenomena are ubiquitous in nature and are essential for the understanding of a variety of processes in biological physics, chemistry and bioengineering~\cite{bird07, schoch08, chowdhury05}.
Prime examples include the paracellular transport of drugs and macromolecules across an epithelium in organs and target-tissues~\cite{panyam03, bareford07}, and the active locomotion of swimming microorganisms in living systems~\cite{bereiter94, ten-hagen11, wang14penetration, liebchen16, bechinger16, menzel15, ruhle17}.

In the microscopic world, fluid motion is well described by the linear Stokes equations, as long as the viscous forces play a dominant role compared to the inertial forces.
In these situations, a full representation of the motion of suspended particles is achieved by the mobility tensor~\cite{kim13, leal80}, which bridges between the velocity moments of the particle and the moments of the force density on its surface.
In biological media, the motion of suspended tracer particles is sensitive to the mechanical state of living cells and the experimentally recorded trajectories can provide useful information about the membrane structure~\cite{sbalzarini05}, or the nature of active processes driving particle motion inside living cells~\cite{gal13}.
Over the last few decades, intracellular particle tracking experiments have widely been utilized as a powerful and often accurate tool for the characterization and diagnostic of individual living cells~\cite{li09, ott13, fodor15, lampo17}, or the determination of cell mechanical properties~\cite{yamada00, chen03, el-kaffas13}.

From a theoretical standpoint, particle motion inside a rigid spherical cavity with fluid velocity satisfying the no-slip boundary condition at the inner cavity is well understood and has been solved since some time ago.
The exact solution of fluid flow takes a particularly simple form when the particle is located at the center of the cavity, and can be determined using the stream function technique, as derived e.g.\ by Happel and Brenner~\cite{happel12}.
The first attempt to obtain the fundamental solution to the Stokes equations due to a point force acting in a Newtonian fluid bounded by a rigid spherical container dates back to Oseen~\cite{oseen28} who used the image solution technique.
Complementary works, which represent extensions of Oseen's solution, commonly known under the name of sphere theorem, have later been presented by Butler~\cite{butler53}, Collins~\cite{collins54, collins58}, Hasimoto~\cite{hasimoto56, hasimoto92, hasimoto97}, Shail~\cite{shail87, shail88}, and Sellier~\cite{sellier08}, to name a few.
A more transparent form of the solution has been presented by Maul and Kim~\cite{maul94, maul96} where both the axisymmetric and asymmetric Stokeslets have been considered independently.
Their results are more useful for computational purposes using boundary integral methods~\cite{pozrikidis01}, and their resolution approach is based on the method presented by Fuentes \textit{et al.}~\cite{fuentes88, fuentes89}.
The latter computed using the image solution technique the flow field due to a Stokeslet acting outside a viscous drop.
The coupling and rotational mobilities have further been reconsidered by Felderhof and Sellier~\cite{felderhof12b}, who employed the point-particle approximation.
The latter is valid when particle radius is very small compared to that of the cavity.
In addition, a combination of multipole expansion and \Faxen's theorem has been used by Zia and collaborators~\citep{aponte16, aponte18}, providing the elements of the grand mobility tensor of finite-sized particles moving inside a rigid spherical cavity. 
Additional works addressed the low-Reynolds-number locomotion inside a viscous drop~\cite{tsemakh04, reigh17, shaik18}, or the dynamics of a particle-encapsulating droplet in flow~\cite{zhan09, zhu17}.

Despite enormous studies on particle motion inside a rigid cavity or a viscous drop, to the best of our knowledge, no works have been yet conducted to investigate particle motion inside a deformable elastic cavity.
Indeed, elastic walls stand apart from rigid boundaries or fluid-fluid interfaces as they endow the system with memory.
Such an effect leads to a long-lived transient anomalous subdiffusive behavior of nearby particles~\cite{daddi16b, daddi16c, rallabandi17, daddi18epje, daddi18jcp}.
Accordingly, particle mobility does not depend only on geometry but also on the forcing frequency of the system.

The goal of this work is to calculate analytically and numerically the frequency-dependent hydrodynamic mobility function of a solid particle slowly moving inside a spherical elastic cavity.
{The membrane cavity is modeled as an infinitely thin (two-dimensional) sheet} made of a hyperelastic material, endowed with shear elasticity and bending rigidity.
Membrane resistance towards shear stresses is modeled by the well-established Skalak model~\cite{skalak73} which is frequently used as a practical model for capsules and red blood cells~\cite{secomb17}.
For calculating the membrane bending forces, we compare two different models.
The first model is based on Helfrich free energy functional~\cite{helfrich73}, often employed for lipid bilayers and biological membranes.
The second model is the linear isotropic model derived from the linear elastic theory of plates and shells~\cite{timoshenko59}.

When the particle is concentric with the cavity, we use the stream function technique to obtain exact solutions of the equations of fluid motion.
For an arbitrary position within the cavity, we use the image solution technique to find analytical expressions of the axisymmetric flow field due to a Stokeslet, in addition to the leading order correction to the particle hydrodynamic mobility function.
Moreover, we investigate the cavity motion, finding that the correction to the pair-mobility function for an arbitrary eccentricity within the cavity is solely determined by membrane shear properties. 
In order to assess the validity and  appropriateness of our analytical predictions, we compare our results with fully-resolved boundary integral simulations where a very good agreement is obtained.

The remainder of the paper is organized as follows.
In Sec.~\ref{sec:theoretical}, we present the stream function technique to obtain exact expressions of the axisymmetric flow field and the hydrodynamic mobility function in the concentric configuration.
We then present in Sec.~\ref{sec:singularity} the image solution technique and compute the particle mobilities in the point-particle framework for arbitrary particle eccentricity within the elastic cavity.
Concluding remarks summarizing our findings and results are contained in Sec.~\ref{sec:conclusions}.


\section{Stream functions}\label{sec:theoretical}

\begin{figure}  
	\centering
	\includegraphics[scale=1]{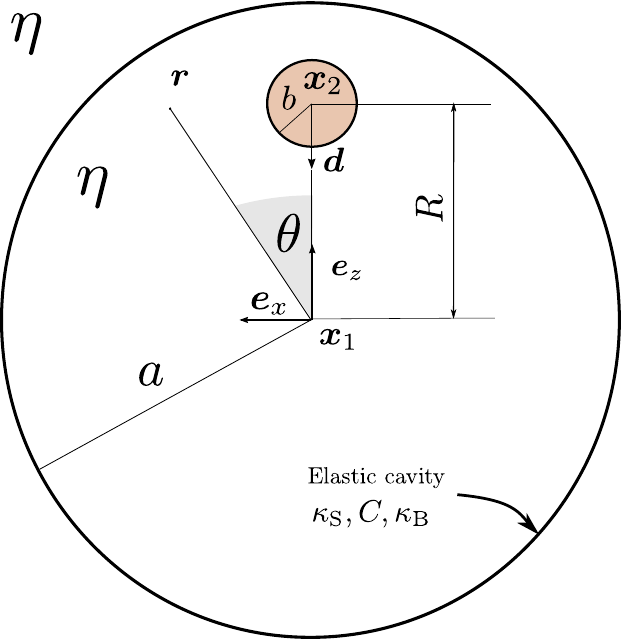}
	\caption{Illustration of the system setup.  
	A small solid particle of radius $b$ positioned at $\xTwo = R \vect{e}_z$ inside a large spherical elastic cavity of radius $a$. The Stokeslet is directed along $\vecD$ connecting the centers of the particle and the cavity.  
	}
	\label{illustration}
\end{figure}

We consider the steady translational motion of a spherical solid particle of radius $b$ inside a large spherical elastic cavity of initial (undeformed) radius $a$.
The origin of coordinates is located at $\xOne$, the center of the cavity, and the solid particle is located at $\xTwo = R \vect{e}_z$, with $R < a-b$, as schematically illustrated in Fig.~\ref{illustration}.

For small amplitude and frequency of motion, the fluid dynamics inside and outside the cavity is governed by the steady Stokes equations
\begin{subequations}
	\begin{align}
	 \eta \boldNabla^2 \vect{v}_\alpha - \boldNabla p_\alpha  &= 0 \, , \label{Stokes:Momentum} \\
	 \boldNabla \cdot \vect{v}_\alpha &= 0 \, , \label{Stokes:Continuity}
	\end{align}
\end{subequations}
where $\alpha = 1$ applies for the fluid on the inside and $\alpha = 2$ for the fluid on the outside.
{Here, we assume that the fluid filling the cavity has the same dynamic viscosity~$\eta$ as the outer fluid.
In real physiological situations, such as, for red blood cells, the viscosity ratio between the interior and exterior fluids is of about five~\citep{zhou95}.
In order to limit the parameter space to be explored, we will, in the present work,  only assess the effect of membrane rigidity on the motion of the encapsulated particle.
The viscosity ratio is expected to have only a quantitative but not a qualitative effect.}

For the sake of convenience, we will scale from now on all the lengths by the cavity radius~$a$.
We begin with the relatively simple situation where the two spheres are concentric.
This corresponds to setting $\xTwo = \xOne$ and $R=0$.
Since, in this situation, the flow is axisymmetric, it is more convenient to express the solution of the equations of motion in term of the stream function.
Accordingly, the solution is reduced to the search of a single scalar function instead of solving simultaneously for the unknown velocity and pressure fields.

The stream functions inside and outside the elastic cavity satisfy~\cite{happel12}
\begin{equation}
 E^4 \psi_\alpha (r,\theta) = 0 \, , \qquad \alpha \in \{1,2\} \, ,  \label{equationForStreamFct}
\end{equation}
where $r$ and $\theta$ are the radial distance and polar angle, respectively, and the operator $E^2$, in spherical coordinates, is given by
\begin{equation}
 E^2 = \frac{\partial^2}{\partial r^2} + \frac{\sin\theta}{r^2} \frac{\partial}{\partial \theta} \left( \frac{1}{\sin\theta} \frac{\partial}{\partial \theta} \right) \, .
\end{equation}

We now assume that the particle moves in the positive~$z$ direction with a constant velocity~$U$.
The no-slip boundary conditions that must be satisfied on the particle surface read~\cite[p.~119]{happel12}
\begin{subequations}\label{no-slip_sphere}
	\begin{align}
	 \psi_1 |_{r=b} &= -\frac{U b^2}{2} \sin^2 \theta \, , \label{no-slip_sphere_1} \\
	 \left. {\psi_1}_{,r} \right|_{r=b}   &= -U b \sin^2 \theta \, . \label{no-slip_sphere_2}
	\end{align}
\end{subequations}
{Additionally, we require the following regularity conditions
\begin{subequations}
	\begin{align}
		|\psi_1| < \infty  &\text{~~as~~} r \to 0 \, ,  \\
		\frac{\psi_2}{r^2} \to 0 &\text{~~as~~} r\to \infty \, .
	\end{align}
\end{subequations}
}

As suggested by the regularity and boundary conditions, the general solution for the steam function in Eq.~\eqref{equationForStreamFct} has previously been derived by Happel and Brenner~\cite{happel12}.
It can be written as
\begin{subequations}
	\begin{align}
	 \psi_1 &= \left( A r + D r^2 + E r^4 + \frac{F}{r} \right) \sin^2 \theta \, , \\
	 \psi_2 &= \left( G r + \frac{H}{r} \right) \sin^2 \theta \, , 
	\end{align}
\end{subequations}
where $A, D, E, F, G,$ and $H$ are six unknown constants to be determined from the boundary conditions imposed at the particle and cavity surfaces.

The flow radial and circumferential velocity components are then computed from the stream functions as
\begin{equation}
 v_r = -\frac{\psi_{,\theta}}{r^2 \sin\theta}  \, , \qquad v_\theta = \frac{\psi_{,r}}{r \sin\theta}  \, ,
\end{equation}
leading to
\begin{subequations}\label{v_1_rThe}
	\begin{align}
	 {v_1}_r        &= -\left( \frac{2A}{r}+2D+2E r^2+\frac{2F}{r^3} \right) \cos \theta \, , \label{v_1_r} \\
	 {v_1}_{\theta} &= \left( \frac{A}{r} + 2D + 4E r^2 - \frac{F}{r^3} \right) \sin \theta \, , \label{v_1_the}
	\end{align}
\end{subequations}
for the fluid on the inside, and 
\begin{subequations}
	\begin{align}
	 {v_2}_r        &= -\frac{2}{r} \left( G + \frac{H}{r^2} \right) \cos \theta \, , \\
	 {v_2}_{\theta} &= \frac{1}{r} \left( G-\frac{H}{r^2} \right) \sin \theta \, , 
	\end{align}
\end{subequations}
for the fluid on the outside.

In addition, the general expressions of the hydrodynamic pressure inside and outside the spherical cavity can readily be determined from the momentum equation \eqref{Stokes:Momentum}, to obtain
\begin{subequations}
	\begin{align}
	 \frac{p_1}{\eta} &= -2\left( \frac{A}{r^2} + 10E r \right) \cos\theta \, , \\
	 \frac{p_2}{\eta} &= -\frac{2G}{r^2} \cos \theta \, .
	\end{align}
\end{subequations}

Having expressed the general solution of fluid motion on both sides of the cavity, we now determine the six unknown coefficients by applying the appropriate boundary conditions: 
(a) the no-slip conditions imposed at the particle surface, given by Eqs.~\eqref{no-slip_sphere} , 
(b) the natural continuity of the fluid velocity between the two sides of the cavity, and 
(c) the discontinuity of the fluid stress tensor due to the presence of the elastic membrane.
Mathematically, we may formulate the problem at hand as
\begin{subequations}\label{BCs}
	\begin{align}
	  [v_r] &= 0 \, , \label{BC:v_r_cont} \\
	  [v_\theta] &= 0 \, , \label{BC:v_theta_cont} \\
	  [\sigma_{\theta r}] &= \Delta f_\theta^{\mathrm{S}} + \Delta f_\theta^{\mathrm{B}} \, , \label{BC:sigma_r_theta} \\
	  [\sigma_{rr}] &= \Delta f_r^{\mathrm{S}} + \Delta f_r^{\mathrm{B}} \, , \label{BC:sigma_r_r} 
	 \end{align}
\end{subequations}
where the notation $[w] := w_2(r=1) - w_1(r=1)$ represents the jump of a quantity $w$ across the membrane.
In spherical coordinates, the non-vanishing components of the fluid stress tensor are expressed in the usual way as~\cite{kim13, dhont96}
\begin{subequations}
	 \begin{align}
	   \sigma_{\theta r} &= \eta \left( v_{\theta,r} - \frac{v_\theta}{r} + \frac{v_{r,\theta}}{r} \right) \, , \label{sigma_r_phi} \\
	   \sigma_{rr} &= -p + 2\eta v_{r,r} \, , \label{sigma_r_r}
	 \end{align}
\end{subequations}
where comma in indices denotes a spatial partial derivative.
Furthermore, $\Delta f_r$ and $\Delta f_\theta$ stand for the radial and circumferential traction jump across the cavity, where the superscripts S and B stand for the shear and bending related parts, respectively.

{
The boundary conditions stated by Eqs.~\eqref{BCs} possess a structure analogous to that of Marangoni flow.
In this context, Tsemakh \textit{et al.}~\cite{tsemakh04} considered the locomotion of a viscous drop encapsulating another smaller drop that serves as a source of a soluble surfactant. 
The tangential stresses at the interface are balanced by interfacial forces due to gradients in the surface tension.
Consequently, the drops undergo an axisymmetric motion driven by Marangoni effect.
It has been demonstrated that, when the drops are not concentric, the inner drop moves toward the surface of the larger drop.
Here, however, we consider driven motion of the encapsulated particle when subjected to an imposed force. 
The tangential and normal hydrodynamic stresses are balanced by the elastic forces, which react back on the surrounding fluid to resist deformation.
Therefore, unlike a surfactant-covered fluid-fluid interface, which, due to concentration gradients and surface tension effects, introduce discontinuities in the stress tensor, elastic membranes stand apart as these discontinuities are intrinsically induced by shear and bending deformation modes. 
}

Continuing, the traction due to shear elasticity according to Skalak model reads
\begin{subequations} \label{tractionJumpShear}
 \begin{align}
  \Delta f_{\theta}^{\mathrm{S}} &= -\frac{2\kS}{3} \bigg( (1+2C) u_{r,\theta} + (1+C) u_{\theta,\theta\theta} \notag \\
		  &+  (1+C)u_{\theta,\theta} \cot\theta -  \left( (1+C) \cot^2 \theta + C \right) u_\theta \bigg)   \, ,   \\
  \Delta f_r^{\mathrm{S}} &=  \frac{2\kS}{3} (1+2C) \left( 2u_r + u_{\theta,\theta} + u_{\theta} \cot \theta \right)   \, , 
 \end{align}
\end{subequations}
where $\kS$ is the shear modulus, and $C$ is a dimensionless number, commonly known in the blood flow community as Skalak parameter~\cite{krueger11, krueger12, barthes16, gekle16, guckenberger18}.
The latter is defined as the ratio between area expansion modulus~$\kA$ and shear modulus~$\kS$.
Moreover, $u_r$ and $u_\theta$ denote the membrane radial and circumferential displacements, respectively.
These are related to the fluid velocity in Fourier space by the no-slip relation imposed at $r=1$.
Specifically~\cite{bickel06}
\begin{equation}
\left.  {v}_\alpha \right|_{r = 1} = i\omega \, {u}_\alpha   \, , \qquad \alpha \in \{r,\theta\} \, .  \label{no-slip}
\end{equation}

Additionally, we include a membrane resistance toward bending, which, can be modeled using the celebrated Helfrich model~\cite{helfrich73, guckenberger16}, or by assuming a linear isotropic model for the bending moments following a thin-shell theory approach~\cite{pozrikidis01jfm}.
The two formulations are equivalent for a planar membrane but not necessarily for membranes of arbitrary geometry~\cite{Guckenberger_preprint}.  
Considering first a linear isotropic model, the traction jumps due to bending are given by~\cite{daddi17b, daddi-thesis}
\begin{subequations}\label{tractionJumpBend}
 \begin{align}
  \Delta f_{\theta}^{\mathrm{B}} &=  {\kB} \bigg( \left(1-\cot^2\theta\right)u_{r,\theta} + u_{r,\theta\theta} \cot\theta + u_{r,\theta\theta\theta} \bigg)   \, , \label{linIsoModTan} \\
  \Delta f_{r}^{\mathrm{B}} &= {\kB} \bigg( \left( 3\cot\theta+\cot^3\theta \right) u_{r,\theta} -  u_{r,\theta\theta}\cot^2\theta  \notag \\
                            &+  2 \cot\theta u_{r,\theta\theta\theta} + u_{r,\theta\theta\theta\theta} \bigg)  \, , \label{linIsoModNormal}
 \end{align}
\end{subequations}
where $\kB$ is the membrane bending modulus.
The traction jump according to Helfrich model reads~\cite{Guckenberger_preprint}
\begin{equation}
\Delta \vect{f}        = -2\kB \left( 2(H^2-K+H_0 H) + \Delta_\parallel \right) (H-H_0) \, \vect{n} \, ,  \label{Paper_Elasticylin_helfrich_tractionJump}
\end{equation}
where $H$ and $K$ are the mean and Gaussian curvatures, respectively, given by
\begin{equation} 
H = \frac{1}{2} \, b_\alpha^\alpha \, , \qquad 
K = \mathrm{det~} b_\alpha^\beta \, , 
\end{equation}
with $b_\alpha^\beta$ being the mixed version of the curvature tensor~\cite{kobayashi63}.
The other quantities are the spontaneous curvature $H_0$, which we take the initial undeformed shape here, the vector normal to the spherical cavity $\vect{n}$, and the Laplace-Beltrami operator $\Delta_{\parallel}$~\cite{deserno15}.
Accordingly, bending introduces a discontinuity only in the normal traction, such that 
\begin{subequations}
	\begin{align}
	   \Delta f_{\theta}^{\mathrm{B}} &= 0 \, , \label{helfrichTan} \\
	   \Delta f_{r}^{\mathrm{B}} &= \kB \bigg( 4u_r+ (5+\cot^2\theta)\cot\theta \,  u_{r,\theta} + (2-\cot^2\theta) u_{r,\theta\theta}  \notag \\
	                             &+ 2\cot\theta \, u_{r,\theta\theta\theta}  + u_{r, \theta\theta\theta\theta} \bigg) \, . \label{helfrichNormal} 
	\end{align}
\end{subequations}
  
It is worth to mention here that the traction jumps due to membrane bending depend only on the normal (radial) displacement.
This behavior is in contrast to the traction jumps due to shear, which, for curved membranes, depend on both the normal and tangential displacements.

Employing the no-slip conditions stated by Eqs.~\eqref{no-slip_sphere} together with the boundary conditions imposed at the membranes, stated by Eqs.~\eqref{BCs}, and solving for the unknown coefficients, we obtain
\begin{subequations}
	\begin{align}
	 A &= -6bU \lambda \left( (1+2C)(1-b^5) \alpha + 5 \right) \, , \\
	 D &=  5bU \lambda \alpha (1+2C) \left( 1-b^2 \right) \, , \\ 
	 E &= -3bU \lambda \alpha (1+2C) (1-b^2) \, , \\
	 F &= 2b^3 U \lambda \left( (1+2C)(1-b^3) \alpha + 5 \right) \, , \\ 
	 G &= A \, , \\
	 H &= 2bU \lambda \left( (1+2C)\left( 1-b^5 \right)\alpha + 5b^2 \right) \, , 
	\end{align}
\end{subequations}
where we have defined 
\begin{equation}
 i\alpha = \frac{2\kS}{3\eta\omega} \, , 
\end{equation}
as a characteristic number associated with membrane resistance towards shear, and 
\begin{equation}
 \lambda^{-1} := 40 + 2\alpha (1-b)(1+2C) \left( 4-b(1+b) \left( 1-9b^2 \right) \right) \, .
\end{equation}
Interestingly, both the linear isotropic and Helfrich bending models lead to analogous expressions of the stream functions.
Therefore, the flow field for a particle concentric with the cavity is solely determined by membrane shear resistance and bending does not play a role.

\subsection{Particle mobility}

The exact analytical solution obtained for the stream functions can be used to assess the effect of the elastic cavity on the motion of a nearby particle, notably for the calculation of the hydrodynamic self-mobility function.

The force exerted by the fluid on the sphere is calculated from the stream function using the formula~\cite[p.~115]{happel12}
\begin{equation}
 F_2 = \eta\pi \int_0^\pi \rho^3 \frac{\partial}{\partial r} \left( \frac{E^2 \psi_1}{\rho^2} \right) r \, \Intd \theta = 8\pi\eta A \, ,  
\end{equation}
which is equivalent to the expression given by Stimson and Jeffery~\cite{stimson26}.
Here $\rho = r\sin\theta$ denotes the polar distance. 
We define the membrane correction factor $K$ as the ratio between the drag in the presence of the outer spherical membrane and the drag in a bulk fluid, such that $F_2 = - 6\pi\eta b U K $.
Then,
\begin{equation}
 K = \frac{4 \left( (1+2C)(1-b^5) \alpha + 5 \right)}{20 + \alpha (1-b)(1+2C) \left( 4-b(1+b) \left( 1-9b^2 \right) \right)} \, .
\end{equation}

Equivalently, the fluid mediated hydrodynamic interactions can also be assessed by determining the correction to the particle self-mobility function, defined in a scaled form as
\begin{equation}
 \frac{ \Delta \mu}{\mu_0}  := \frac{1}{K} - 1 = - \frac{5}{4} \frac{b\alpha (1+2C)\left(1-b^2\right)^2}{ 5 + \alpha (1+2C)(1-b^5) } \, , \label{mobiCorrCentered}
\end{equation}
where $\mu_0 = 1/(6\pi\eta b)$ is the usual bulk mobility.
Not surprisingly, the frequency-dependent particle mobility is solely determined by membrane shear properties.
At leading order in $b$, Eq.~\eqref{mobiCorrCentered} can be expanded as 
\begin{equation}
 \frac{ \Delta \mu}{\mu_0} = - \frac{5}{4} \frac{\alpha(1+2C)}{ 5+\alpha(1+2C) } \, b + \bigO (b^3) \, , \label{mobiCorrCentered_leadingOrder}
\end{equation}
and is commonly denominated the mobility correction in the point-particle approximation.
Taking $\alpha\to\infty$, corresponding to an infinite shear modulus, or equivalently, to a vanishing frequency, we obtain
\begin{equation}
 \lim_{\alpha\to\infty} \frac{ \Delta \mu}{\mu_0} = -\frac{5}{4} \frac{b\left(1-b^2\right)^2}{1-b^5} = -\frac{5}{4} \, b + \bigO (b^3) \, . \label{mobiCorrCentered_vanishingFreq}
\end{equation}

\begin{figure}
 \centering
 \includegraphics[scale=1]{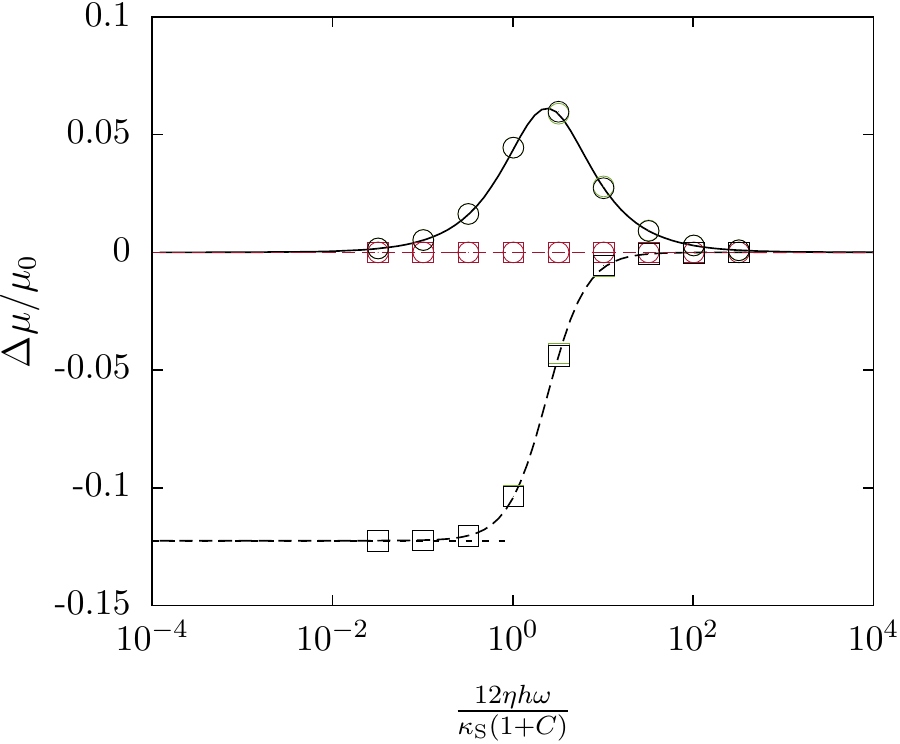}
 \caption{(Color online) The scaled self-mobility correction versus the scaled frequency~$\beta$ inside a spherical elastic cavity whose membrane is endowed with pure shear (green), pure bending (red) of both shear and bending rigidities (black). The small solid particle has a radius $b=1/10$, concentric with a large spherical cavity of unit radius. 
 For the membrane parameters, we take $C=1$ and a reduced bending modulus $\EB=8/3$.
 The analytical predictions are shown as solid lines for the imaginary parts and as dashed lines for the real parts.
 {Since the self mobility is solely determined by membrane shear resistance, the green and black lines are indistinguishable.}
 Symbols refer to the boundary integral simulations where squares are for the real part and circles are for the imaginary part. 
 The horizontal dashed line represent the vanishing frequency limit predicted by Eq.~\eqref{mobiCorrCentered_vanishingFreq}.
 }
 \label{deltaMuPart_centered}
\end{figure}

For a rigid spherical cavity with no-slip boundary conditions at the inner surface, the cavity does not move and thus creating an additional resistance to the motion of the particle.
Accordingly, the particle mobility is obtained as
\begin{equation}
  \mu_{\mathrm{R}} = \lim_{\alpha\to\infty} \mu - \frac{1}{6\pi\eta} \, , \label{rigidVSvanishingFreq}
\end{equation}
where the subscript R stands for rigid, and the term subtracted on the right hand side is the bulk mobility of the cavity.
Scaling by the particle bulk mobility, the correction for a hard cavity reads
\begin{equation}
 \frac{ \Delta \mu_{\mathrm{R}}}{\mu_0} = - b \left( 1 + \frac{5}{4} \frac{\left(1-b^2\right)^2}{1-b^5} \right) = -\frac{9}{4} \, b + \bigO (b^3) \, . \label{mobilityCorrection_hard}
\end{equation}

The latter result is in full agreement with the solution by Happel and Brenner~\cite{happel12}, and with the solution by Aponte-Rivera and Zia~\cite{aponte16}, who accounted for the particle finite-size up to the 5th order in $b$.
Therefore, apart from a term $b$, the mobility in the vanishing frequency limit for an elastic cavity, as given by Eq.~\eqref{mobiCorrCentered_vanishingFreq}, is identical to that obtained inside a rigid cavity given by Eq.~\eqref{mobilityCorrection_hard}.
Indeed, the additional term is due to the fact that the rigid cavity remains at rest while the elastic cavity necessarily undergoes translational motion.

In a way analogous to a planar elastic membrane~\cite{daddi16}, we define the characteristic frequency for shear as $\beta := 6B\eta h\omega/\kS$ where $B:=2/(1+C)$, and $h = 1-R$ is the distance from the particle center to the closest point on the cavity surface, such that $h=1$ for concentric spheres. 
In Fig.~\ref{deltaMuPart_centered}, we show the scaled correction to the frequency-dependent self mobility versus the scaled frequency~$\beta$.
Here the particle has a radius $b=1/10$, concentric with a spherical elastic cavity of unit radius.
We consider the situations where the membrane is endowed with pure shear (green), pure bending (red), or both rigidities (black).
We take a Skalak parameter $C=1$ and a reduced bending modulus $\EB:=\kB/(h^2 \kS) = 8/3$.

We observe that the correction to the particle mobility depends uniquely on membrane shear resistance, and thus in full agreement  with our theoretical calculations.
The real part of the hydrodynamic mobility correction (shown as dashed line) is a logistic-like function whereas the imaginary part exhibits at intermediate frequencies around $\beta\sim 1$ the typical peak structure.
The latter is a clear signature of the memory effect induced by the elastic nature of the membrane on the system.
In the high frequency limit, the correction to the mobility vanish and thus the behavior in a bulk fluid is recovered.
In the low frequency limit, the correction approaches that predicted theoretically by Eq.~\eqref{mobiCorrCentered_vanishingFreq}, being the same, apart from a term $b$, as the hard cavity limit given by Eq.~\eqref{mobilityCorrection_hard}. 
A prefect agreement is obtained between the exact analytical calculations and the numerical simulations we have performed using a completed double layer boundary integral method.

\subsection{Cavity motion}

In the following, we examine the motion of the cavity induced by a concentric solid particle translating along the $z$ direction.
For that purpose, we define the pair-mobility function $\mu^{12}$ as the ratio between the capsule velocity $V_1$ and the force exerted by the solid particle on the fluid.
The net translational velocity of the cavity can be computed by volume integration of the $z$ component of the fluid velocity inside the cavity.
Specifically~\cite{felderhof06b}
\begin{equation}
 V_1 (\omega) = \frac{2\pi}{\Omega} \int_0^{\pi} \int_b^1 {v_1}_z (r,\theta, \omega) \, r^2 \sin \theta \,  \Intd r \, \Intd \theta \, ,  \label{cavityVelocity}
\end{equation}
where $\Omega := 4\pi/3$ is the volume of the undeformed cavity and ${v_1}_z = {v_1}_r \cos \theta - {v_1}_{\theta} \sin \theta$ is the fluid velocity along the $z$~direction.
The radial and circumferential velocities are given by Eqs.~\eqref{v_1_rThe}.
This leads to the pair-mobility function, written in a scaled form as
\begin{equation}
 \begin{split}
  6\pi\eta \mu^{12} &= \frac{3}{2} \left( 1-b^2 \right) \notag \\
                    &-\frac{\alpha}{4} \frac{(1+2C)(1-b)^3(1+b) \left( 3b^2(2+b) + 2(1+2b) \right)}{5 + \alpha (1+2C) \left( 1-b^5 \right)} \, .  \label{pairMobi_centered}
 \end{split}
\end{equation}
The first term on the right-hand side of the latter equation represents the bulk contribution stemming from the Stokeslet solution in an unbounded medium, whereas the second term is the frequency-dependent correction due to the presence of the elastic cavity.
The correction can therefore be expressed as a Debye-like model with a single relaxation time, given by
\begin{equation}
 \tau = \frac{15}{2(1+2C)\left( 1-b^5 \right)} \frac{\eta}{\kS} \, . \label{Debye}
\end{equation}

\begin{figure}
 \centering
 \includegraphics[scale=1]{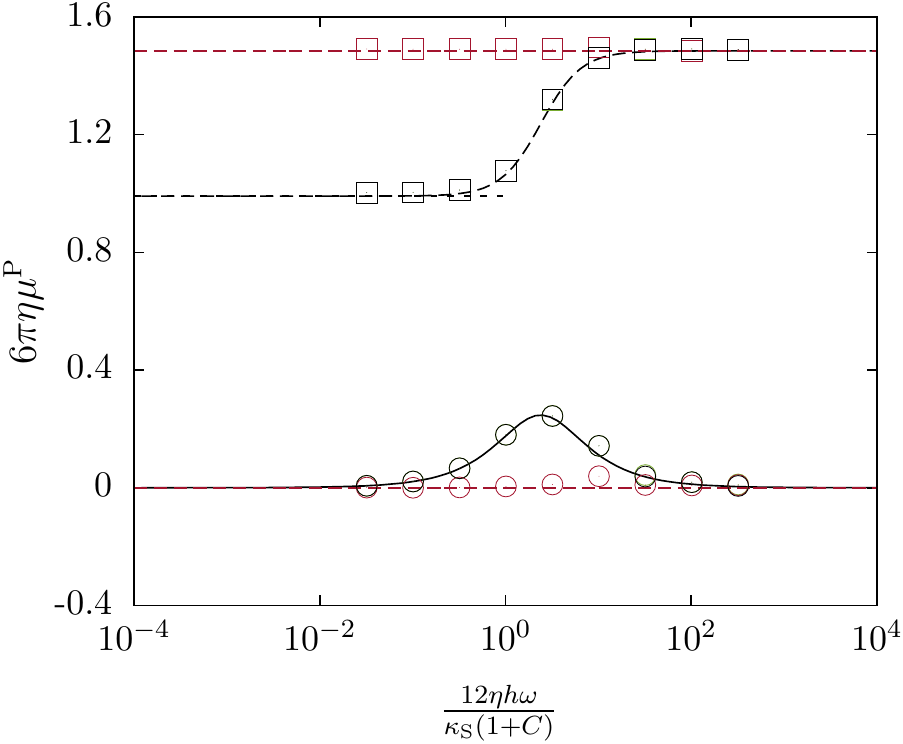}
 \caption{(Color online) {The pair-mobility function scaled by $1/(6\pi\eta)$} versus the scaled frequency for a membrane cavity endowed with pure shear (green), pure bending (red) or both rigidities (black).
 The analytical prediction given by Eq.~\eqref{pairMobi_centered} is shown as dashed and solid lines for the real and imaginary parts, respectively.
 Symbols refer to the corresponding BIM simulations. 
 Horizontal dashed in the low frequency regime corresponds to the limit predicted theoretically by Eq.~\eqref{pairMobi_centered_vanishingFreq}.
 Here we use the same particle/membrane parameters as in Fig.~\ref{deltaMuPart_centered}.}
 \label{deltaMuCaps_centered}
\end{figure}

At leading order in $b$, the scaled pair-mobility given by Eq.~\eqref{pairMobi_centered} reads
\begin{equation}
 6\pi\eta \mu^{12} = \frac{3}{2} - \frac{\alpha}{2} \frac{1+2C}{5 + \alpha (1+2C)} + \bigO (b^2) \, . \label{pairMobi_centered_leadingOrder}
\end{equation}
Taking $\alpha\to\infty$, Eq.~\eqref{pairMobi_centered} yields 
\begin{equation}
 \lim_{\alpha\to\infty} 6\pi\eta \mu^{12} = \frac{(1-b^2) \left( 4 + b^3 \left( 5-9b^2 \right) \right)}{4 \left( 1-b^5 \right)} = 1 + \bigO \left( b^2 \right) \, . \label{pairMobi_centered_vanishingFreq}
\end{equation}

We further mention that the hydrodynamic force acting by the fluid on the cavity internal surface~$S_1$ is readily determined by integrating the normal stress vector over the surface, to obtain
\begin{equation}
 \F_1 = -\int_{S_1} \boldSigma \cdot \eR \, \Intd S = -8\pi\eta A \, \vect{e}_z  \, ,
\end{equation}
which is found to be the same in magnitude but opposite in sign as the force $\F_2$ acting by the fluid on the solid particle.

In Fig.~\ref{deltaMuCaps_centered} we show the scaled pair-mobility function versus the scaled frequency~$\beta$ using the same parameters as in Fig.~\ref{deltaMuPart_centered}.
For a membrane with pure bending, the real part of the pair-mobility amounts to $(3/2)(1-b^2)$ while the imaginary part vanishes, corresponding to the behavior in a bulk fluid. 
In contrast to that, a membrane endowed with shear resistance shows a richer dynamics, where the pair mobility depends strongly on the actuation frequency.
Indeed, the pair mobility for a membrane possessing both shear and bending rigidities is undistinguished from that of a membrane with a pure shear.
An excellent agreement is obtained between the analytical theory and boundary integral simulations.

Analogous exact analytical predictions using the stream function technique can in principle be carried out for any arbitrary position within the spherical cavity.
The general solution may conveniently be expressed in term of an infinite series involving Legendre polynomials~\cite{happel12}.
Nevertheless, due the complex nature of the underlying boundary conditions, the resolution is laborious and non-trivial.
In order to overcome this difficulty, we will employ, as an alternative way, a fundamentally different approach based on the image solution technique to compute the flow field induced by a Stokeslet acting inside a spherical elastic cavity.
This will result to the computation of the hydrodynamic mobility function in the point-particle approximation, valid when $b \ll 1$, as is detailed in the next section.

\section{Singularity solution}\label{sec:singularity}

The following image solution technique has originally been proposed by Fuentes \textit{et al.}~\cite{fuentes88, fuentes89}, who computed the flow field induced by a point force acting outside a spherical drop.
The same approach has been employed by some of us in earlier works~\cite{daddi17b, daddi17c} to address the fluid motion induced by a point force acting nearby a spherical elastic membrane with shear and bending rigidities.
It is worth mentioning that this method has recently been employed by Shaik and Ardekani to calculate the Stokeslet singularity outside a drop covered with an incompressible surfactant~\cite{shaik17}.

The fluid flow inside the cavity can be written as a sum of two distinct contributions
\begin{equation}
 \vect{v}_1 = \vStok + \vect{v}^* \, , 
\end{equation}
where, $\vStok$ is the velocity field induced by a point force acting at the particle position $\xTwo$, and $\vect{v}^*$ is the image solution required {to satisfy the boundary conditions stated by Eqs.~\eqref{BCs}, in addition to the regularity conditions:
\begin{subequations}\label{regularitySingularitySolution}
	\begin{align}
		|\vect{v}_1| < \infty &\text{~~as~~} r\to 0 \, , \\
		\vect{v}_2 \to \mathbf{0} &\text{~~as~~} r\to \infty \, .
	\end{align}
\end{subequations}
}

Now we briefly sketch the main resolution steps.
First, the velocity $\vStok$ due to the Stokeslet is written in terms of spherical harmonics, which are then transformed into harmonics based at~$\xOne$ via the Legendre expansion.
Second, the image system solution $\vect{v}^{*}$ and the solution outside the cavity~$\vect{v}_2$ are, respectively, expressed as interior and exterior harmonics based at~$\xOne$ using Lamb's general solution~\cite{lamb32, cox69}.
The last step consists of determining the series unknown expansion coefficients by satisfying the boundary conditions at the membrane surface, given by Eqs.~\eqref{BCs}.
{We note that the hydrodynamic fields are expanded in terms of harmonics such that the regularity conditions stated by Eqs.~\eqref{regularitySingularitySolution} are satisfied.}

\subsection{Stokeslet solution}

We begin with writing the Stokeslet acting at $\xTwo$,
\begin{equation}
 \vStokcom_i =  \G_{ij} F_j = \frac{1}{8\pi\eta} \left( F_i \frac{1}{s} + F_j (\vect{x} - \xTwo)_i {\nabla_2}_j \frac{1}{s} \right) \, , 
 \label{stokeslet_at_X2}
\end{equation}
where $s:= |\vect{x}-\xTwo|$.
Here ${\nabla_2}_j := {\partial }/{\partial  {x_2}_j}$ stands for the nabla operator taken with respect to the singularity position~$\xTwo$.
Using Legendre expansion, the harmonics based at $\xTwo$ can be expanded as
\begin{equation}
 \frac{1}{s} = \infSum R^n \, \varphi_n (r, \theta)  \, , 
\end{equation}
where the unit vector $\vecD := ( \xOne - \xTwo)/R = -\vect{e}_z$, the position vector $\R = \X - \xOne$, and $r := |\R|$.
Furthermore, $\varphi_n$ are the harmonics of degree~$n$, related to the Legendre polynomials of degree~$n$ by~\cite{abramowitz72}
\begin{equation}
 \varphi_n (r, \theta) := \frac{(\vecD \cdot \boldNabla)^n}{n!} \frac{1}{r} = \frac{1}{r^{n+1}} \, P_n (\cos \theta) \, .
\end{equation}

For the axisymmetric case, the force is exerted along the unit vector $\vecD$, and can thus be written as $\vect{F} = F \vecD$.
By making use of the identities
\begin{equation}
 \boldNabla_2 R^n = -nR^{n-1} \, \vecD \, , \quad  (\vecD \cdot \boldNabla_2)\, \vecD = 0 \, ,  \label{twoIdentities}
\end{equation}
Eq.~\eqref{stokeslet_at_X2} therefore becomes
\begin{equation}
 \begin{split}
  \vStok &= -\frac{F}{8\pi\eta} \infSum R^{n-1} \big[ (n-1)R \, \vecD + n \, \R \big] \varphi_n \, .
  \label{vStok_withDependentHarmonics}
 \end{split}	
\end{equation}
Hence, the Stokeslet at $\xTwo$ is written in terms of harmonics based at $\xOne$.
Note that the terms with $\vecD\,\varphi_n$ in Eq.~\eqref{vStok_withDependentHarmonics} are not independent harmonics.
For their elimination, we will use~\cite{daddi17b}
\begin{equation}
 \begin{split}
  \vecD\,\varphi_n &= \frac{1}{2n+1} \bigg[ \boldNabla \varphi_{n-1} - r^2 \boldNabla \varphi_{n+1} 
                   -(2n+3) \R \, \varphi_{n+1} \bigg] \, ,  \notag 
 \end{split}
\end{equation}
leading after substitution into Eq.~\eqref{vStok_withDependentHarmonics} to
\begin{align}
 \vStok &= \frac{F}{8\pi\eta} \infSumOne \bigg[ \left( \frac{n-2}{2n-1}\, r^2 R^{n-1} - \frac{n}{2n+3} \, R^{n+1} \right) \boldNabla \varphi_n \notag  \\
        &-\frac{2(n+1)}{2n-1} \, R^{n-1} \vect{r} \varphi_n \bigg] \, .
\end{align}
For future reference, we will state explicitly the projected velocity components onto the radial (normal) and circumferential (tangential) directions.
For that purpose, we will make use of the following identities 
\begin{subequations}
 \begin{align}
  \eR \cdot \boldNabla \varphi_n &= -\frac{n+1}{r} \, \varphi_n \, , \label{projectionRad_1} \\
  \eR \cdot \R \varphi_n &= r \varphi_n \, , \label{projectionRad_2} \\
  \eThe \cdot \R \varphi_n &= 0 \, , \label{projectionTan}
 \end{align}
\end{subequations}
leading to the final expression of the Stokeslet solution
\begin{align}
 \vStokcom_r &= \frac{F}{8\pi\eta} \infSumOne \left[ -\frac{n(n+1)}{2n-1}\, r R^{n-1} + \frac{n(n+1)}{2n+3} \frac{R^{n+1}}{r} \right] \varphi_n \, , \notag \\
 \vStokcom_\theta &= \frac{F}{8\pi\eta} \infSumOne \left( \frac{n-2}{2n-1}\, r^2 R^{n-1} - \frac{n}{2n+3} \, R^{n+1} \right) \psi_n \, , \notag
\end{align}
where we have defined 
\begin{equation}
 \psi_n := \eThe \cdot \boldNabla \varphi_n   = \frac{1}{r} \frac{\partial \varphi_n}{\partial \theta} \, . \label{psi_n}
\end{equation}
The pressure can directly be determined from the integration of Eq.~\eqref{Stokes:Momentum}, to obtain
\begin{equation}
 \pStok = \frac{F}{8\pi} \infSumOne -2n R^{n-1} \varphi_n \, .
\end{equation}

We further note that $\varphi_n$ and $\psi_n$ constitute sets of independent harmonics satisfying the properties
\begin{subequations}
	\begin{align}
	 \int_{0}^{\pi} \varphi_n \varphi_m \sin\theta \, \Intd \theta &= \frac{2}{2n+1} \frac{\delta_{mn}}{r^{2n+2}} \, . \label{orthoProp_phi} \\
	 \int_{0}^{\pi} \psi_{m}  \psi_{n}  \sin\theta \, \Intd \theta &= \frac{2n(n+1)}{2n+1} \frac{\delta_{mn}}{r^{2n+4}} \, . \label{orthoProp_psi}
	\end{align}
\end{subequations}

In the following, the image system solution in addition to the solution inside the cavity will be derived.

\subsection{Image system solution}
For the image system solution inside the cavity, we use Lamb's general solution~\cite{cox69, misbah06}, which can be written in terms of \emph{interior} harmonics based at $\xOne$ as 
\begin{equation}
 \begin{split}
  \vect{v}^{*} &= \frac{F}{8\pi\eta} \sum_{n=1}^{\infty} \Bigg[ 
                  A_n \bigg( \frac{n+3}{2}r^{2n+3} \, \boldNabla \varphi_n  \\
                  &+ \frac{(n+1)(2n+3)}{2} \, r^{2n+1} \R \varphi_n \bigg) \\
                  &+ B_n \left( r^{2n+1} \boldNabla \varphi_n + (2n+1)r^{2n-1} \R \varphi_n \right)
		  \Bigg] \, . \label{imageSysSolution}
 \end{split}
\end{equation}

After making use of Eqs.~\eqref{projectionRad_1} through \eqref{projectionTan}, the projected components of the image system solution read
\begin{subequations}
	\begin{align}
	 v_r^* &= \frac{F}{8\pi\eta} \infSumOne \left[ \frac{n(n+1)}{2} \, r^{2n+2} A_n + nr^{2n} B_n \right]  \varphi_n \, , \\
	 v_\theta^* &= \frac{F}{8\pi\eta} \infSumOne \left( \frac{n+3}{2} \, r^{2n+3} A_n + r^{2n+1} B_n \right)  \psi_n \, ,
	\end{align}
\end{subequations}
while the solution for the pressure field inside the cavity is obtained as
\begin{equation}
 p^* = \frac{F}{8\pi} \infSumOne (n+1)(2n+3) A_n r^{2n+1} \varphi_n \, .
\end{equation}

\subsection{Solution outside the cavity}
We use Lamb's general solution which can be written in terms of \emph{exterior} harmonics based at $\xOne$ as 
\begin{equation}
 \begin{split}
  \vect{v}_2 &= \frac{F}{8\pi\eta} \sum_{n=1}^{\infty} \Bigg[ 
                  a_n \bigg( -\frac{n-2}{2} \, r^2 \, \boldNabla \varphi_n 
                  + (n+1) \, \R \varphi_n \bigg) \\
                  &+ b_n \boldNabla \varphi_n \, , 
		  \Bigg] \, ,
 \end{split}
\end{equation}
which can be projected onto normal and tangential components, to obtain
\begin{subequations}
	\begin{align}
	 {v_2}_r      &= \frac{F}{8\pi\eta} \infSumOne \left[ \frac{n(n+1)}{2} \,r a_n - \frac{n+1}{r} \, b_n \right] \varphi_n \, , \\
	 {v_2}_\theta &= \frac{F}{8\pi\eta} \infSumOne \left( -\frac{n-2}{2} \, r^2 a_n + b_n \right) \psi_n \, .
	\end{align}
\end{subequations}
Lastly, the pressure field outside the cavity can then be presented as
\begin{equation}
 p_2 = \frac{F}{8\pi} \infSumOne n(2n-1) a_n \varphi_n \, .
\end{equation}

Having expressed the general solution for the velocity and pressure fields, we now proceed for the determination of the unknown coefficients inside the cavity $A_n$ and $B_n$, and outside the cavity $a_n$ and $b_n$.

\subsection{Determination of the series coefficients}

\subsubsection{Continuity of velocity}

The continuity of the radial and circumferential velocities as stated by Eqs.~\eqref{BC:v_r_cont} and~\eqref{BC:v_theta_cont} leads to
\begin{align}
  \frac{a_n}{2} - \frac{b_n}{n} &= \frac{A_n}{2} + \frac{B_n}{n+1} -\frac{R^{n-1}}{2n-1} + \frac{R^{n+1}}{2n+3} \, , \notag  \\
 -\frac{n-2}{2}\, a_n+b_n &= \frac{n+3}{2} \, A_n + B_n + \frac{n-2}{2n-1}\, R^{n-1} - \frac{n R^{n+1}}{2n+3}  \, . \notag 
\end{align}
Solving these two equations for $a_n$ and $b_n$, the coefficients outside the cavity can be expressed in terms of those inside as 
\begin{subequations}\label{Eq_1und2}
	\begin{align}
	 a_n &=  \frac{2n+3}{2}\, A_n + \frac{2n+1}{n+1} \, B_n -\frac{2}{2n-1} \, R^{n-1} \, , \label{Eq_1} \\
	 b_n &=  \frac{n(2n+1)}{4} \, A_n + \frac{n(2n-1)}{2(n+1)} \, B_n -\frac{n}{2n+3} \, R^{n+1} \, . \label{Eq_2}
	\end{align}
\end{subequations}

The coefficients $A_n$ and $B_n$ can be determined from the traction jump equations stemming from membrane shear and bending resistances.
In order to probe the effect of these two elasticity modes in more depth, we will consider in the following idealized membranes with pure shear or pure bending resistances.

\subsubsection{Discontinuity of stress tensor}

\paragraph{Shear contribution}

We first consider a membrane with only-shear resistance, such as that of a typical artificial capsule designed for drug delivery~\cite{Freund_2014, zhu14, barthes16}.
It follows from Eqs.~\eqref{BCs}, representing the tangential and normal traction jumps, that
\begin{subequations}
	\begin{align}
	    \left[ v_{\theta,r}\right] &= 
	     -\alpha \bigg( (1+2C) v_{r,\theta} + (1+C) \left( v_{\theta,\theta\theta} + v_{\theta,\theta} \cot\theta \right)  \notag \\
	            &- \left.   \left( (1+C) \cot^2 \theta + C \right) v_\theta \bigg) \right|_{r=1}  \, , \label{jump_v_Phi_Shear} \\
	    \bigg[ -\frac{p}{\eta} \bigg]  &= 
	    {\left. -\alpha (1+2C) v_{r,r} \right|_{r=1}} \, , \label{jump_v_r_Shear}
	\end{align}
\end{subequations}
where, again, $i\alpha := 2\kS/(3\eta\omega)$ is the shear parameter. 
In order to handle the derivatives with respect to $r$, we will make use of the identities
\begin{align}
 \varphi_{n,r} = -\frac{n+1}{r} \, \varphi_n \, , \qquad  \psi_{n,r} = -\frac{n+2}{r} \, \psi_n \, .
\end{align}

By making use of the orthogonality property given by Eq.~\eqref{orthoProp_psi}, together with
\begin{equation}
  \begin{split}
    & \int_{0}^{\pi} \psi_{m} \left(\psi_{n,\theta\theta} + \psi_{n,\theta} \cot \theta - \psi_{n} \cot^2 \theta \right) \sin\theta \, \Intd \theta \\
               &=  -\frac{2n(n+1)(n^2+n-1)}{2n+1} \frac{\delta_{mn}}{r^{2n+4}} \, , \label{orthoProp_psi_2}
  \end{split}
\end{equation}
the tangential traction jump equation given by Eq.~\eqref{jump_v_Phi_Shear} leads to 
\begin{equation}
 \begin{split}
  (2n+1) & (2n+3) A_n + \frac{2(4n^2-1)}{n+1} \, B_n \\
   &= 
  \alpha \bigg( (1+2C) (n+1)\left( {n} a_n- 2b_n \right)  \\ 
  &+ \left( 1-(1+C)n(n+1) \right) \left( -{(n-2)} a_n + 2b_n \right) \bigg) \, . \label{Eq_3_She}
 \end{split}
\end{equation}

Further, the shear-related contribution to the normal traction equation, given by Eq.~\eqref{jump_v_r_Shear} results to
\begin{equation}
 \begin{split}
 (n-2) & (2n+1)(2n+3) A_n + \frac{2n(4n^2-1)}{n+1} \, B_n  \\
 &= \alpha (1+2C) (n+1) \left( -{n^2} a_n + 2(n+2) b_n \right) \, . \label{Eq_4_She}
 \end{split}
\end{equation}

Eqs.~\eqref{Eq_3_She} and~\eqref{Eq_4_She} together with Eqs.~\eqref{Eq_1und2} form a closed system of linear equations, amenable to direct resolution via the standard substitution technique.
We obtain
\begin{subequations}
	\begin{align}
	 A_n &= \frac{\alpha n(n+2)}{K} \left( \frac{2n+1}{2n+3} \, R^{n+1} \, K_{+} -  R^{n-1} \, K_{-} \right) \, , \label{An_She} \\
	 B_n &= \frac{\alpha (n+1)}{2M} \left( \frac{2n+1}{2n-1} \, R^{n-1} \, M_{-} - n(n+2) R^{n+1} \, M_{+} \right) \, ,  \label{Bn_She}
	\end{align}
\end{subequations}
where the coefficients $K$, $K_{\pm}$, $M$ and $M_{\pm}$ have rather complex and lengthy expressions and are therefore moved to the Appendix.
Particularly, by considering $\alpha \to \infty$, corresponding to taking an infinite shear modulus, or, a vanishing frequency limit, we obtain
\begin{subequations}\label{AnBn_alphaInfty}
	\begin{align}
	 \lim_{\alpha\to\infty} A_n &= \frac{2n+1}{2n+3} \, R^{n+1} - R^{n-1} \, , \label{An_alphaInfty} \\
	 \lim_{\alpha\to\infty} B_n &= \frac{(n+1)(2n+1)}{2(2n-1)} \, R^{n-1} -\frac{n+1}{2} \, R^{n+1} \, . \label{Bn_alphaInfty}
	\end{align}
\end{subequations}
The latter limits correspond to the solution obtained for a point force acting inside a rigid spherical cavity with no-slip boundary conditions.
Moreover, both $a_n$ and $b_n$ vanish in this limit in which the fluid outside the cavity is at rest.

\vspace{0.33cm}
\paragraph{Bending contribution}

Next, we consider a membrane endowed with pure bending resistance such as that of a fluid vesicle~\cite{noguchi05, kaoui12, kaoui16, nait-ouhra18}.
As already pointed out, two models are commonly used to describe membrane resistance towards bending.
We will first provide explicit analytical expressions by assuming a linear isotropic model for the bending moments.
The corresponding traction jump equations given by Eqs.~\eqref{BC:sigma_r_theta} and~\eqref{BC:sigma_r_r} read
\begin{subequations}
	\begin{align}
	  \left[ v_{\theta,r} \right]   &= \left. \alphaB \bigg( \left(1-\cot^2\theta\right)v_{r,\theta} + v_{r,\theta\theta} \cot\theta + v_{r,\theta\theta\theta} \bigg) \right|_{r=1} \, , \label{jump_v_Phi_Bending} \\
	  \bigg[ -\frac{p}{\eta} \bigg] &=  \alphaB \bigg( \left( 3\cot\theta+\cot^3\theta \right) v_{r,\theta} -  \cot^2 \theta \, v_{r,\theta\theta}  \notag \\
	                                &+ \left. 2 \cot\theta \, v_{r,\theta\theta\theta}+ v_{r,\theta\theta\theta\theta} \bigg) \right|_{r=1} \, , \label{jump_v_R_Bending}
	 \end{align}
\end{subequations}
where, $i \alphaB := \kB /(\eta \omega)$ is the bending parameter.
By making use of Eqs.~\eqref{orthoProp_psi} and~\eqref{orthoProp_psi_2}, the tangential traction jump reads
\begin{equation}
 \begin{split}
  (2n+1) & (2n+3)A_n + \frac{2(4n^2-1)}{n+1} \, B_n \\
         &= \alphaB (n^2+n-2) (n+1)\left( n a_n - 2 b_n \right) \, ,  \label{Eq_3_Ben_LIS}
 \end{split}
\end{equation}

Continuing, using Eq.~\eqref{orthoProp_phi} together with the orthogonality relation 
\begin{equation}
 \begin{split}
  &\int_{0}^{\pi} \varphi_m \bigg( \left( 3\cot\theta+\cot^3\theta \right) \varphi_{n,\theta} -  \varphi_{n,\theta\theta}\cot^2\theta 
      + 2 \varphi_{n,\theta\theta\theta}\cot\theta \\
      &+ \varphi_{n,\theta\theta\theta\theta} \bigg) \sin\theta \Intd \theta 
      = \frac{2n(n-1)(n+1)(n+2)}{2n+1} \frac{\delta_{mn}}{r^{2n+2}} \, , 
 \end{split}
\end{equation}
the normal traction jump reads
\begin{equation}
 \begin{split}
  (n-2) & (2n+1)(2n+3) A_n + \frac{2n(4n^2-1)}{n+1} \, B_n  \\
        &= -\alphaB n(n-1)(n+1)^2(n+2) \left( na_n - 2b_n \right) \, . \label{Eq_4_Ben_LIS}
 \end{split}
\end{equation}

Solving the system of linear equations arising from Eqs.~\eqref{Eq_1und2} together with \eqref{Eq_3_Ben_LIS} and~\eqref{Eq_4_Ben_LIS} leads to the determination of the unknown coefficients.
We obtain
\begin{subequations}
	\begin{align}
	 A_n &= \frac{\alphaB n^2(n+2)^2 (n^2-1)}{K_\mathrm{B}} \left( \frac{2n-1}{2n+3}\, R^{n+1} - R^{n-1} \right) \, , \\
	 B_n &= \frac{\alphaB n(n+2)(n-1)(n+1)^2 (n^2+2n-2)}{2K_\mathrm{B}}   \notag \\
	     &\times \left(\frac{2n+3}{2n-1} \, R^{n-1} -  R^{n+1} \right) \, ,  \label{Bn_Ben_LIS}
	\end{align}
\end{subequations}
where
\begin{equation}
 \begin{split}
  K_\mathrm{B} &= 2\alphaB n^6+6\alphaB n^5-\alphaB n^4+4(2-3\alphaB)n^3 \\
               &+ (12-\alphaB)n^2 + 2(3\alphaB-1)n - 3 \, .
 \end{split}
\end{equation}
By taking the limit $\alphaB \to \infty$, which corresponds to taking an infinite membrane bending modulus, or a vanishing forcing frequency, the two coefficients read
\begin{align}
 \lim_{\alphaB\to\infty} A_n &= \frac{n(n+2)}{2n^2+2n-3} \left( \frac{2n+3}{2n-1} \, R^{n-1} -  R^{n+1} \right) \, , \notag  \\
 \lim_{\alphaB\to\infty} B_n &= \frac{(n+1)(n^2+2n-2)}{2(2n^2+2n-3)} \left(\frac{2n+3}{2n-1} \, R^{n-1} -  R^{n+1} \right) \, , \notag  
\end{align}
which are found to be different from the solution previously obtained when taking $\alpha\to\infty$ in a shear-only membrane, as can be seen from Eqs.~\eqref{AnBn_alphaInfty}.

We next consider Helfrich model for membrane bending, which leads to the traction jumps equations
\begin{subequations}\label{tanNorTractionHelfrich}
	\begin{align}
	 [v_{\theta, r}] &= 0 \, , \label{tanTractionHelfrich} \\
	 ~\bigg[ -\frac{p}{\eta} \bigg] &= \alphaB \bigg( 4v_r+ (5+\cot^2\theta)\cot\theta \,  v_{r,\theta} \notag \\
	                                &+  \left. (2-\cot^2\theta) v_{r,\theta\theta} + 2\cot\theta \, v_{r,\theta\theta\theta}  + v_{r, \theta\theta\theta\theta} \bigg) \right|_{r=1} \, . \label{norTractionHelfrich}
	\end{align}
\end{subequations}

After making use of the orthogonality property \eqref{orthoProp_phi}, together with
\begin{equation}
 \begin{split}
  \int_0^\pi & \varphi_m \big( 4\varphi_n+ (5+\cot^2\theta)\cot\theta \,  \varphi_{n,\theta} + (2-\cot^2\theta) \varphi_{n,\theta\theta}  \\
             &+ 2\cot\theta \, \varphi_{n,\theta\theta\theta}  + \varphi_{n, \theta\theta\theta\theta} \big) \sin \theta \, \Intd \theta \\
             &= \frac{2(n+2)^2 (n-1)^2}{2n+1} \frac{\delta_{mn}}{r^{2n+2}} \, , \notag
 \end{split}
\end{equation}
we obtain the two following equations
	\begin{align}
	 (2n+1) & (2n+3)A_n + \frac{2(4n^2-1)}{n+1} \, B_n = 0 \, , \notag \\
	 (n-2)  & (2n+1)(2n+3)A_n + \frac{2n(4n^2-1)}{n+1}\, B_n  \notag \\
	        &= -\alphaB (n+2)^2 (n-1)^2 (n+1) (na_n-2b_n) \, , \notag
	\end{align}
which, upon making use of Eqs.~\eqref{Eq_1und2}, and solving for the coefficients $A_n$ and $B_n$, yields
\begin{subequations}
	\begin{align}
	 A_n &= \frac{\alphaB (n-1)^2 n (n+1) (n+2)^2}{K_\mathrm{H}} \left( \frac{2n-1}{2n+3}\, R^{n+1} - R^{n-1} \right) \, , \\
	 B_n &= \frac{\alphaB (n-1)^2 n (n+1)^2 (n+2)^2}{2K_\mathrm{H}} \left( \frac{2n+3}{2n-1}\, R^{n-1} - R^{n+1} \right) \, , \label{Bn_Ben_Helfrich}
	\end{align}
\end{subequations}
where
\begin{equation}
 \begin{split}
  K_\mathrm{H} &= 2\alphaB n^6+6\alphaB n^5-2\alphaB n^4 + 2(4-7\alphaB)n^3 \\
               &+ 12n^2 + 2(4\alpha-1)n - 3 \, .
 \end{split}
\end{equation}
Similarly, by taking the limit $\alphaB \to \infty$, Eqs.~\eqref{tanNorTractionHelfrich} leads to
\begin{subequations}
	\begin{align}
	 \lim_{\alphaB\to\infty} A_n &= \frac{1}{2} \left( \frac{2n-1}{2n+3}\, R^{n+1} - R^{n-1} \right) \, , \label{An_alphaInfty_Helfrich} \\
	 \lim_{\alphaB\to\infty} B_n &= \frac{n+1}{4}  \left( \frac{2n+3}{2n-1}\, R^{n-1} - R^{n+1} \right) \, . \label{Bn_alphaInfty_Helfrich}
	\end{align}
\end{subequations}
Clearly, these coefficients also differ from those obtained previously for a shear-only membrane given by Eqs.~\eqref{AnBn_alphaInfty}.

\subsection{Particle mobility}

The leading-order particle mobility is obtained by evaluating the image system solution given by Eq.~\eqref{imageSysSolution} at the particle position as
\begin{equation}
 \left. \vect{v}^* \right|_{\X = \xTwo} = \Delta \mu \, \F \, ,  
\end{equation}
leading to the particle mobility correction, which can conveniently be written in a scaled form as an infinite series 
\begin{equation}
 \frac{\Delta\mu}{\mu_0} = -\frac{3b}{8} \infSumOne \left[ {n(n+1)} R^{n+1} A_n + 2n R^{n-1} B_n \right] \, . \label{mobilityCorrection}
\end{equation}

In the particular case of $R=0$, which corresponds to the concentric case earlier considered in Sec.~\ref{sec:theoretical}, only the term with $n=1$ remains.
We thus recover the leading-order self mobility 
\begin{equation}
 \left. \frac{ \Delta \mu}{\mu_0} \right|_{R=0} = -\frac{3b}{4} \, B_1 = - \frac{5}{4} \frac{\alpha(1+2C)}{ 5+\alpha(1+2C) } \, b  \, , 
\end{equation}
in full agreement with Eq.~\eqref{mobiCorrCentered_leadingOrder}, obtained using the stream function technique.
Clearly, the mobility correction depends only on membrane resistance towards shear since $B_1=0$ for bending-only membranes (see Eqs.~\eqref{Bn_Ben_LIS} and~\eqref{Bn_Ben_Helfrich} for the general expressions of $B_n$ using the two bending models.)

Now, by taking the limit $\alpha\to\infty$ in Eq.~\eqref{mobilityCorrection}, the correction to the particle self mobility reads
\begin{equation}
 \lim_{\alpha\to\infty} \frac{\Delta\mu}{\mu_0} = b \left( 1-\frac{9}{4} \frac{1}{1-R^2} \right) \, . \label{mobilityCorrectionVanishingFreq}
\end{equation}
The same limit is obtained when considering a membrane with pure shear.
For a large cavity radius, Eq.~\eqref{mobilityCorrectionVanishingFreq} reduces to the leading-order mobility correction near a no-slip planar wall, as first obtained using the method of reflection by Lorentz~\cite{lorentz07}.
Specifically,
\begin{equation}
  \lim_{\alpha\to\infty} \frac{\Delta\mu}{\mu_0}  = -\frac{9}{8} \frac{b}{h} + \bigO \left( \frac{1}{a} \right) \, .
\end{equation}

By considering the coefficients \eqref{An_alphaInfty} and~\eqref{Bn_alphaInfty} associated with a hard cavity, the scaled correction to the particle mobility reads~\cite{sellier08, felderhof12b}
\begin{equation}
 \frac{\Delta\mu_{\mathrm{R}}}{\mu_0} = -\frac{9}{4} \frac{b}{1-R^2} \, , \label{mobilityCorrectionHard}
\end{equation}
being identical to the leading-order correction given for $R=0$ by Eq.~\eqref{mobilityCorrection_hard}.
Again, the mobility inside a hard cavity is recovered in the vanishing frequency limit apart from a term~$b$, as explained by Eq.~\eqref{rigidVSvanishingFreq}.

In fact, the sum over $n$ in Eq.~\eqref{mobilityCorrection} and the limit when $\alpha\to\infty$ cannot be swapped.
In other words, taking the limit when $\alpha\to\infty$ before evaluating the sum as is the case for a hard cavity does not lead to the same result as evaluating the sum first and then taking the limit, as is done for an elastic cavity.
This is justified by the fact that the dominated convergence theorem does not hold here for the infinite series given by Eq.~\eqref{mobilityCorrection}.

Now, by considering a membrane with pure bending resistance modeled by Helfrich model, the mobility correction in the vanishing frequency limit reads
\begin{equation}
  \lim_{\alphaB\to\infty} \frac{\Delta\mu}{\mu_0} = b \left(1-\frac{15}{8} \frac{1}{1-R^2} \right) \, . \label{mobilityCorrectionVanishingFreq_Bending}
\end{equation}
We further recover for large cavity radius the well-known mobility correction near a planar interface separating two fluids having the same viscosity, namely~\cite{lee79, lee80, bickel07}
\begin{equation}
 \lim_{\alphaB\to\infty} \frac{\Delta\mu}{\mu_0}  = -\frac{15}{16} \frac{b}{h} + \bigO \left( \frac{1}{a} \right) \, .
\end{equation}

\begin{figure}
 \centering
 \includegraphics[scale=1]{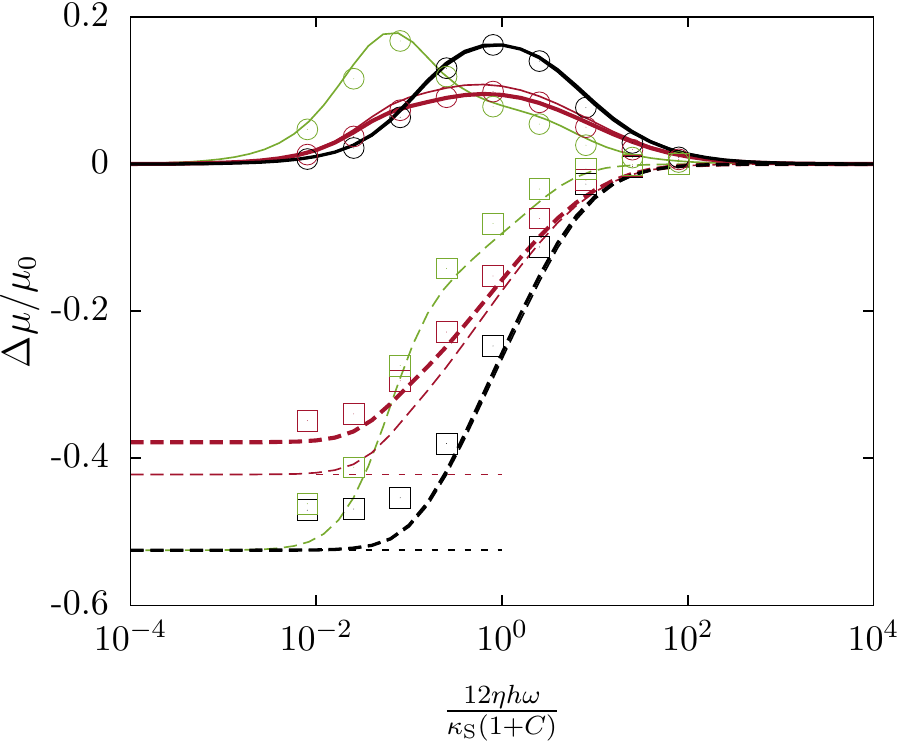}
 \caption{(Color online) The scaled frequency-dependent mobility correction versus the scale frequency inside a spherical elastic cavity with pure shear (green), pure bending (red) and both shear and bending (black).
 The thin and thick red lines correspond to the linear isotropic model and Helfrich model respectively.
 The particle has a radius $b=1/10$ and is positioned at $R=4/5$.   
 Horizontal dashes lines shown in black and red correspond to the vanishing-frequency limits predicted by Eqs.~\eqref{mobilityCorrectionVanishingFreq} and \eqref{mobilityCorrectionVanishingFreq_Bending}, respectively.
 Here we use the same membrane parameters as in Fig.~\ref{deltaMuPart_centered}.
 }
 \label{deltaMu_OFFcenter}
\end{figure}

In Fig.~\ref{deltaMu_OFFcenter}, we show the scaled frequency-dependent self-mobility correction versus the scaled frequency~$\beta$ for a particle of radius $b=1/10$ located at $R=4/5$ inside a spherical cavity.
Unlike the situation where the particle is concentric with the cavity, a contribution from bending resistance arises. 
We observe that Helfrich model (thick red lines) leads to a better agreement with the BIM simulations than the linear isotropic model (thin red lines).  
Considering the shear-only membrane, we observe that a second peak of more pronounced amplitude arises in the low frequency regime.
This peak does not occur in planar membranes but has been observed previously for a particle moving outside a large spherical capsule~\cite{daddi17b, daddi17c}.
In fact, the peak is attributed to the fact that the traction jumps due to shear involve a contribution from the normal displacement, in contrast to planar membranes, where these traction jumps depend solely on the in-plane tangential displacements.
Only one single peak however occurs for a bending-only membrane for both models since the traction jumps due to bending involve only the normal deformations and thus explaining the absence of the second peak.

{
It is worth noting that the calculation of the hydrodynamic mobility function based upon the singularity solution is valid only in the far-field limit, i.e.\ when $h \gg b$.
For separation distances $h \sim b$, the far-field approximation falls apart.
In this situation, it becomes  necessary to consider higher-order reflections.
However, even though $h$ is taken only one particle diameter, i.e.\ $h=2b$, the point-particle approximation is remarkably found to yield a good prediction of the particle mobilities when compared with fully-resolved BIM simulations.}

\subsection{Cavity motion}

\begin{figure}
 \centering
 \includegraphics[scale=1]{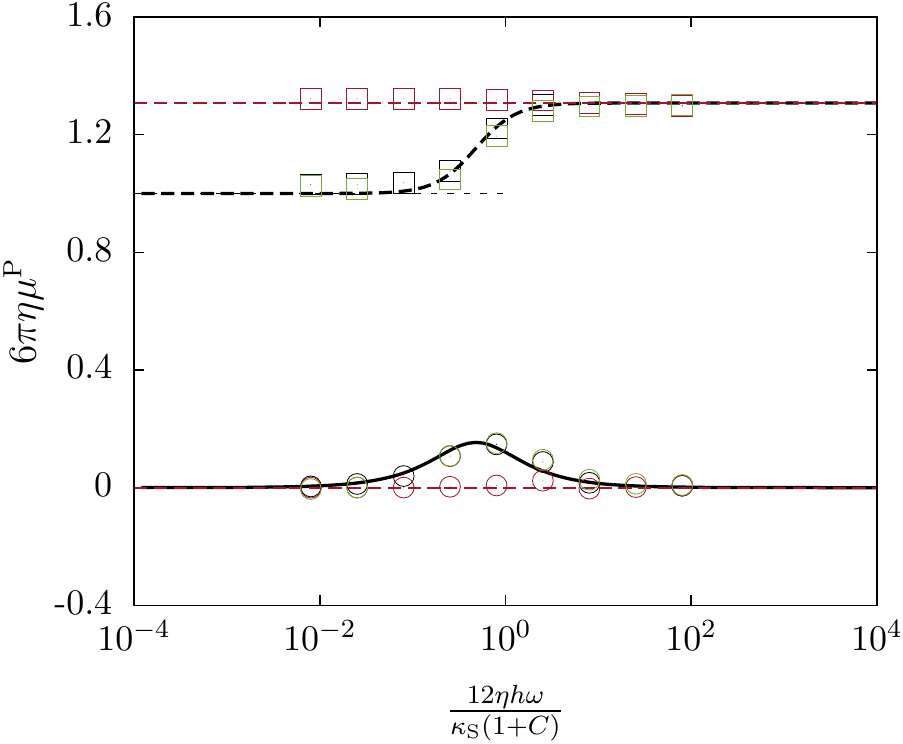}
 \caption{(Color online) {The pair-mobility function scaled by $1/(6\pi\eta)$} versus the scaled frequency for cavity with only-shear (green), only-bending (red) or both shear and bending (black).
 The particle has a radius $b=1/10$ and is positioned at $R=4/5$. 
 The analytical prediction stated by Eq.~\eqref{pairMobi_offCenter} is shown as dashed and solid lines for the real and imaginary parts, respectively, while symbols are the BIM simulations. 
 Horizontal dashed in the low frequency regime corresponds to one.
 For the membrane parameters, see Fig.~\ref{deltaMuPart_centered}.}
 \label{deltaMu_OFFcenter_capsMotion}
\end{figure}

Finally, the cavity translational velocity is computed by integrating the fluid velocity as stated by Eq.~\eqref{cavityVelocity}, with the exception that the radial variable $r$ is now integrated between 0 and~1.
We find that only the term with $n=1$ of the series remains, leading to
\begin{equation}
 \mu^{12} = -\frac{1}{8\pi\eta} \left( A_1 + B_1 - 2 + \frac{2}{5} \, R^2 \right) \, , 
\end{equation}
which, upon substitution of the coefficients with their expressions, yields
\begin{equation}
 6\pi\eta \mu^{12} = \frac{3}{2} -\frac{3}{10}\, R^2 - \frac{5-3R^2}{10} \frac{\alpha (1+2C)}{5 + \alpha (1+2C)}   \, . \label{pairMobi_offCenter}
\end{equation}

Interestingly, even for $R \ne 0$, the pair mobility depends solely on membrane shear.
Therefore, bending does not play a role, i.e.\ in a way similar to that previously observed for two concentric spheres.
As $\alpha\to\infty$, the pair mobility tends to unity, independently of the value of $R$.
In particular, for $R=0$ we recover the leading-order solution given by Eq.~\eqref{pairMobi_centered_leadingOrder}, obtained using the stream function technique.
The correction to the pair mobility follows a Debye-like model with a relaxation time given by the leading-order term in Eq.~\eqref{Debye}.

Similarly, it can be shown that the force exerted by the fluid on the internal surface of the cavity is equal in magnitude but opposite in sign to the friction force $\F_2$ acting on the particle.

In Fig.~\ref{deltaMu_OFFcenter_capsMotion} we show the scaled pair-mobility function versus the scaled frequency using the same parameters as in Fig.~\ref{deltaMu_OFFcenter}.
The pair mobility for a bending-only membrane remains unchanged and amounts to $3/2 - 3R^2/10$ in the whole range of forcing frequencies. 
For a cavity with a finite shear resistance, the real part is a monotonically increasing function of frequency that varies between 1 and $3/2 - 3R^2/10$, while the imaginary parts exhibits the usual bell-shaped behavior with the typical peak at $\beta \sim 1$.
Our analytical predictions are favorably compared with BIM simulations.

{
Due to the motion of the encapsulated particle, the imbalance in the hydrodynamic stress tensor across the membranes leads to cavity deformation.
The latter can be determined from the flow velocity field via the no-slip condition imposed at $r=1$.
In Fourier space, this condition reads
\begin{equation}
	\vect{u}(\theta) = \left. \frac{\vect{v}_1 (r,\theta)}{i\omega} \right|_{r=1} 
			 = \left. \frac{\vect{v}_2 (r,\theta)}{i\omega} \right|_{r=1} \, .
			 \label{displacementCavity}
\end{equation}
}

{
Neglecting the bending contribution, and assuming that the shear parameter~$|\alpha| \gg 1$, the series coefficients $a_n$ and $b_n$ scale as $\alpha^{-1} \sim i\omega\eta / \kappa_\mathrm{S}$.
Therefore, the fluid velocity field $v_2 \sim i\omega F / \kappa_\mathrm{S}$.
It follows from Eq.~\eqref{displacementCavity} that the membrane displacement field $u \sim F / \kappa_\mathrm{S}$.
As a result, the deformation of the cavity becomes important when $F/\kappa_\mathrm{S} \sim h$.
In the present work, we have considered a force amplitude~$F \ll \kappa_\mathrm{S} h$ such that the membrane deformation remains sufficiently small.}

\section{Conclusions}\label{sec:conclusions}

In this paper, we have presented a fully analytical theory of the low-Reynolds number motion of small particle slowly moving inside a large spherical elastic cavity. 
We have modeled the membrane resistance towards shear forces by Skalak model which incorporates into a single strain energy functional both the resistance towards shear and area conservation.
We have assessed two different models for bending namely Helfrich model and the linear isotropic model.

We have first solved the underlying equations of fluid motion in the relatively simple scenario, where the particle is concentric with the large spherical cavity. 
In this situation, exact analytical solutions are obtained and expressed in a closed mathematical form using the stream function technique.
We have found that the solution of the flow problem is solely determined by membrane shear and that bending does not play a role. 
Moreover, we have shown that, in the vanishing frequency limit, the particle hydrodynamic mobility is higher than that obtained inside a rigid cavity with no-slip boundary conditions at its inner surface.
This behavior has been justified by the fact that a steady rigid cavity exerts an additional hindrance in particle motion, reducing particle hydrodynamic mobility in a significant way.

For an arbitrary position of the particle within the spherical cavity, we have employed the image solution technique to find analytical expression of the axisymmetric flow field induced by a point force acting on the fluid domain.
This lead to expressions of the mobility function in the point-particle framework, valid when the particle size is smaller than that of the spherical elastic cavity.
Considering the motion of the cavity, we have found that the pair-mobility function depends uniquely on membrane shear properties.
This behavior has been shown to be true for any arbitrary value of particle eccentricity.
For example setups, we have favorably compared our analytical predictions with fully resolved numerical simulations performed using a completed double layer boundary integral method. 

{
The solution presented in the present work is limited to the axisymmetric situation in which the velocity field is decomposed onto radial and circumferential components.
The asymmetric situation in which the particle is moving tangent to the spherical cavity can be explored in future studies.
In conjunction with the results obtained here, the motion induced by external forces directed along an arbitrary direction, can thus, in this way, be determined.}


\begin{acknowledgments}
 We thank Maciej Lisicki for helpful discussions.
 Funding from the DFG (Deutsche Forschungsgemeinschaft) within DA~2107/1-1 (ADMI) and LO~418/19-1 (HL) is gratefully acknowledged. 
 SG thanks the Volkswagen Foundation for financial support and acknowledges the Gauss Center for Supercomputing e.V. for providing computing time on the GCS Supercomputer SuperMUC at Leibniz Supercomputing Center. 
 We acknowledge support from the COST Action MP1305, supported by COST (European Cooperation in Science and Technology).
\end{acknowledgments}

\section*{Contributors}

ADMI conceived the study, performed the analytical calculations and numerical simulations. ADMI, HL, and SG interpreted the results and drafted the manuscript.
All authors discussed the results and approved the manuscript.

\begin{widetext}

\appendix*

\section{Mathematical expression}

The analytical expressions of the function appearing in Eqs.~\eqref{An_She} and~\eqref{Bn_She} are given for the series coefficient $A_n$ by
\begin{align}
 K_{+} &= 2(1+C)n^3 + \left( (2\alpha+5)C+\alpha+1 \right)n^2 + (1+C)n - (1+2C)(1+\alpha) \, , \notag \\
 K_{-} &= 2(1+C)n^3 + \left( (2\alpha+3)C+\alpha-1 \right)n^2 + (1+C)n - (1+2C)\alpha +1 \, , \notag \\
 K     &= 4(1+C)\alpha n^5 + \left( 16+(1+2C)\alpha^2+10(1+C)\alpha \right)n^4 + 2\left( 16+(1+2C)\alpha^2+3(1+C)\alpha \right)n^3 \notag \\
       &+ \left( 8-(1+2C)\alpha^2-(1+C)\alpha \right)n^2 -\left( 8+2(1+2C)\alpha^2 + (7+C)\alpha \right)n - 3(1+\alpha) \, , \notag
\end{align}
and for $B_n$ by
\begin{align}
 M_{-} &= -12+2(1+C)n^5 + \left( (2\alpha+5)C+\alpha+1 \right)n^4+ \left( (4\alpha+3)C+2\alpha-5 \right)n^3 + \left( 9-(1+2C)\alpha \right)n^2 \notag \\
       &-2 \left( (1+2C)\alpha-5 \right)n \, , \notag \\
 M_{+} &= 2(1+C)n^3 + \left( (2\alpha+3)C+\alpha-1 \right) n^2 + (1+C)n - \alpha(1+2C) + 1 \, , \notag \\
 M     &= 4(1+C)\alpha n^5 + \left( 16+(1+2C)\alpha^2+10 (1+C) \alpha \right) n^4 + 2 \left( 16+(1+2C)\alpha^2+3 (1+C) \alpha \right) n^3 \notag \\
       &+ \left( 8-(1+2C)\alpha^2- (1+C) \alpha \right) n^2 -\left( 8+2(1+2C) \alpha^2 + (C+7) \alpha \right)n -3(1+\alpha) \, , \notag
\end{align}
\end{widetext}

%

\end{document}